\documentstyle[12pt,aps,psfig]{revtex}
\begin{document}

\def\be{\begin{equation}}
\def\ee{\end{equation}}
\def\ba{\begin{eqnarray}}
\def\ea{\end{eqnarray}}

\title{Mean--Field Approximation to the Master Equation for
  Sympathetic Cooling of Trapped Bosons}
\author{S. J. Wang$^1$, M. C. Nemes$^2$, A. N. Salgueiro$^3$, and
  H. A. Weidenm\"uller$^3$}
\address{$^1$ Department of Modern Physics, Lanzhou University,
  730000 Lanzhou, P.R. China}
\address{$^2$Department of Physics, University of Belo Horizonte, Belo
  Horizonte, Minas Gerais, Brasil}
\address{$^3$Max-Planck-Institut f\"ur Kernphysik, D-69029 Heidelberg,
  Germany}

\date{\today}

\maketitle

\begin{abstract}We use the mean--field approximation to simplify
the master equation for sympathetic cooling of Bosons. For the
mean single--particle occupation numbers, this approach yields the
same equations as the factorization assumption introduced in an
erlier paper. The stationary or equilibrium solution of the
resulting master equation for the one--body density matrix shows
that the mean--field approximation breaks down whenever the
fraction of condensate Bosons exceeds ten percent or so of the
total. Using group--theoretical methods, we also solve the
time--dependent master equation for the one--body density matrix.
Given the time dependence of the mean single--particle occupation
numbers, this solution is obtained by quadratures. It tends
asymptotically towards the equilibrium solution.

\end{abstract}

\section{Introduction}
\label{int}

Sympathetic cooling of Bosons or Fermions is an important tool in
Atom Optics. It is used in cases where the interaction between
atoms is too weak for evaporative cooling to work. This fact calls
for the development of a theory describing the process. A master
equation for sympathetic cooling was formulated in Ref.~\cite{lew95}.
This equation turned out to be too unwieldy for practical purposes.
A significant simplification was introduced in Refs.~\cite{pap01}.
Two independent approximation schemes -- microcanonical averaging,
and a factorization assumption -- both led to equations that could
be solved numerically. The results were in good agreement with each
other~\cite{pap01}.

In this paper, we introduce and analyze another approximation to
the master equation of Ref.~\cite{lew95}. This is the mean--field
approximation often used in many--body theory. We show that this
approach leads to the same rate equation for the mean occupation
numbers as the factorization assumption introduced in
Refs.~\cite{pap01}. In this sense, the mean--field approach contains
and generalizes the factorization assumption. It allows us to study
not only the time evolution of the mean occupation numbers but
that of the full one--body density matrix. We do so by using a
group--theoretical technique in von--Neumann space. We are able to
construct analytically both the equilibrium and the non--equilibrium
solutions of the master equation for the one--body density matrix.
This leads to a complete theoretical understanding of equilibrium
and non--equilibrium properties of the solutions of the master
equation and, at the same time, yields insight into the limitations
of the mean--field approach itself. We focus attention on the case
of Bosons which is theoretically more complex and, therefore, more
interesting.

In Section~\ref{mas} we formulate the master equation and use
decoherence to reduce it to diagonal form. The mean--field
approximation is introduced in Section~\ref{mea}. Our algebraic
technique is introduced in Section~\ref{alg}. We then investigate
the equilibrium solution (Section~\ref{sta}). While this
investigation could have been done without using the algebra
developed in Section~\ref{alg}, our method shows its full power
when we diagonalize the rate operator in Section~\ref{dia} and when
we construct the time--dependent solutions of the master equation
in Section~\ref{tim}. Section~\ref{sum} contains a brief summary.

\section{The Master Equation for the Density Matrix and Decoherence}
\label{mas}

We follow the notation of Refs.~\cite{lew95,pap01}. System $A$ is
subject to sympathetic cooling and consists of $N_A$ Bosons. The
master equation for the time--dependence of the reduced density
matrix $\rho_A(t)$ has the form
\begin{equation}
\label{eq1}
\frac{{\rm d} \rho_A(t)}{{\rm d}t} = - \frac{i}{\hbar} \biggl [ H_A +
H'_{A-A}, \rho(t) \biggr ] + {\cal L} \rho_A \ .
\end{equation}
Here, $H_A$ is the sum of the single--particle Hamiltonians for the
atoms in system $A$ containing the kinetic energy and the
harmonic--oscillator potential of the trap, while $H'_{A-A}$ contains
the interaction between the atoms in system $A$. The action of the
Liouvillean $\cal L$ on the reduced density matrix $\rho_A(t)$ is
given by
\begin{eqnarray}
\label{eq2}
{\cal L} \rho_A &=& \sum_{{\vec n}, {\vec n}', {\vec m}, {\vec m}'}
\Gamma^{{\vec m},{\vec m}'}_{{\vec n},{\vec n}'} \biggl ( 2
a^{\dagger}_{\vec m} a_{{\vec m}'} \rho_A(t) a^{\dagger}_{\vec n}
a_{{\vec n}'} - a^{\dagger}_{\vec n} a_{{\vec n}'} a^{\dagger}_{\vec
  m} a_{{\vec m}'} \rho_A(t) \nonumber \\
&&\qquad \qquad \qquad \qquad \qquad - \rho_A(t) a^{\dagger}_{\vec n}
a_{{\vec n}'} a^{\dagger}_{\vec m} a_{{\vec m}'} \biggr ) \ .
\end{eqnarray}
The rate coefficients $\Gamma^{{\vec m},{\vec m}'}_{{\vec n},{\vec
    n}'}$ are given in Ref.~\cite{lew95}; we do not repeat the
definition here. Suffice it to say that they account for the
interaction between particles in system $A$ and those in the cooling
system $B$. The latter is assumed to be at thermal equilibrium at all
times. The labels ${\vec m}$ and ${\vec n}$ refer to the three quantum
numbers $(m_x,m_y,m_z)$ defining the single--particle eigenstates of
an isotropic harmonic oscillator in three dimensions representing the
trap potential, with associated creation and annihilation operators
$a_{\vec m}^{\dagger}$ and $a_{\vec m}$, respectively.
Eqs.~(\ref{eq1}) and (\ref{eq2}) are derived from the von Neumann
equation for the density matrix of the combined system $A$ plus $B$
under the following assumptions: The process is Markovian, the
correlation time for the interaction between systems $A$ and $B$ is
much shorter than the cooling time, and a rotating--wave approximation
applies. Equations of the form of Eqs.~(\ref{eq1},\ref{eq2}) are often
used to discuss Boson condensation phenomena~\cite{scu99,koc00}.

It was shown in Ref.~\cite{pap01} that if the number $N_B$ of
particles in the bath is very large, decoherence acts very quickly and
reduces the density matrix to diagonal form. Then, Eq.~(\ref{eq2})
takes the form
\ba
{\cal L} \rho_A &=& \sum_{{\vec m} \neq {\vec n}}
\Gamma^{{\vec m},{\vec n}}_{{\vec n},{\vec m}} \biggl ( 2
a^{\dagger}_{\vec m} a_{{\vec n}} \rho_A(t) a^{\dagger}_{\vec n}
a_{{\vec m}} - a^{\dagger}_{\vec n} a_{{\vec m}} a^{\dagger}_{\vec
  m} a_{{\vec n}} \rho_A(t) \nonumber \\
&&\qquad \qquad \qquad \qquad \qquad - \rho_A(t) a^{\dagger}_{\vec n}
a_{{\vec m}} a^{\dagger}_{\vec m} a_{{\vec n}} \biggr ) \ .
\label{eq3}
\ea

\section{Mean--Field Approximation}
\label{mea}

Eq.~(\ref{eq3}) serves as the starting point for our mean--field
approximation. We apply this approximation in standard fashion by
replacing on the right--hand side of Eq.~(\ref{eq3}) one pair of
creation and annihilation operators referring to the same
single--particle state ${\vec m}$ or ${\vec n}$ by its expectation
value. We use the notation
\be
\label{eq3a}
\langle a_{\vec m}^{\dagger} a_{\vec m} \rangle = {\rm tr} ( a_{\vec
  m}^{\dagger} a_{\vec m} \rho_A ) = N_{\vec m} \ .
\ee
The quantities $N_{\vec m}$ are the mean occupation numbers of the
orbital ${\vec m}$. For the term ${\cal L} \rho_A$, this procedure
yields
\ba
\label{eq4}
{\cal L} \rho_A &=& \sum_{{\vec m} \neq {\vec n}} \Biggl (
\Gamma^{{\vec m},{\vec n}}_{{\vec n},{\vec m}} N_{\vec n} \biggl ( 2
a^{\dagger}_{\vec m} \rho_A(t) a_{\vec m} -  a^{\dagger}_{\vec
  m} a_{\vec m} \rho_A(t) - \rho_A(t) a_{\vec m} a^{\dagger}_{\vec m}
- \rho_A \biggr ) \nonumber \\
&&\qquad \qquad \qquad + \Gamma^{{\vec n}{\vec
    m}}_{{\vec m}{\vec n}} (N_{\vec n} + 1) \biggl ( 2
a_{\vec m} \rho_A(t) a^{\dagger}_{\vec m} -  a^{\dagger}_{\vec m}
a_{\vec m} \rho_A(t) - \rho_A(t) a_{\vec m} a^{\dagger}_{\vec m}
+ \rho_A \biggr ) \Biggr ) \ .
\ea
Eqs.(\ref{eq1},\ref{eq3a}) and (\ref{eq4}) imply
\be
\label{eq5}
\frac{{\rm d} {\rm tr} \rho_A(t)}{{\rm d}t} = 0 \ .
\ee
In what follows, we explore the algebraic structure of Eq.~(\ref{eq4}).
We show that this equation is linear in the generators of the group
SU(1,1). In the algebraic theory of non--autonomous quantum
systems~\cite{wan93,wan89}, it is shown that any such linear dynamical
system is integrable and, thus, can be solved analytically.

\section{Algebraic Structure of the Mean--Field Equation}
\label{alg}

In the first two Subsections, we introduce the mathematical tools
which are used in Subsection~\ref{str} to explore the structure of
Eq.~(\ref{eq4}).

\subsection{Right and Left Algebras}
\label{rig}

Let $| N \rangle_{\vec m}$ stand for the $N$--fold occupied
single--particle state ${\vec m}$. In a Hilbert--space basis
defined in terms of such states, the creation and annihilation
operators $a^{\dagger}_{\vec m}$ and $a_{\vec m}$ appearing in
Eq.~(\ref{eq4}) may operate from the left on the ket state $| N
\rangle_{\vec m}$ or from the right on the bra state ${_{\vec m}}
\langle N |$. In the latter case, we attach an upper index $r$ to the
operators and write them to the left of the density matrix
$\rho_A$. We do so with the understanding that for a product of $m$
operators $O_1 \times O_2 \times \ldots \times O_m$, we have
\be
\label{eq6c}
O^r_1 \times O^r_2 \times \ldots \times O^r_m \rho_A = \rho_A O_m
\times \ldots \times O_2 \times O_1 \ .
\ee
This convention (the reversal of order as we bring the factors from
one side of $\rho_A$ to the other) is the cause for the differences in
signs between the first and the second set of Eqs.~(\ref{eq7}) below.
As a result, the pair of operators $a^{\dagger}_{\vec m}$ and $a_{\vec
  m}$ generates both the left and and the right representations of
the group hw(4). We define accordingly
\be
\label{eq6a}
a^{l \dagger}_{\vec m} = a^{\dagger}_{\vec m} \ ,
a^{l}_{\vec m} = a_{\vec m} \ ,
n^{l}_{\vec m} = a^{\dagger}_{\vec m} a_{\vec m} \ ,
\ee
if the operators act from the left on the ket state $| N \rangle_{\vec
  m}$, and
\be
\label{eq6b}
a^{r \dagger}_{\vec m} = a^{\dagger}_{\vec m} \ , \nonumber \\
a^{r}_{\vec m} = a_{\vec m} \ , \nonumber \\
n^{r}_{\vec m} = a^{\dagger}_{\vec m} a_{\vec m} \ ,
\ee
if the operators act from the right on the bra state ${_{\vec m}}
\langle N |$. Together with the unit operator, the three operators
appearing on the left--hand sides of Eqs.~(\ref{eq6a}) (of
Eqs.~(\ref{eq6b})) constitute the left representation hw(4)$_l$ (the
right representation hw(4)$_r$ of hw(4), respectively). These
operators obey the following commutation relations.
\ba
\label{eq7}
&&hw(4)_l: [n^l_{\vec m}, a^{l, \dagger}_{\vec m}] = + a^{l,
  \dagger}_{\vec m} \ , \nonumber \\
&&\qquad \qquad [n^l_{\vec m}, a^{l}_{\vec m}] = - a^{l}_{\vec m} \ ,
  \nonumber \\
&&\qquad \qquad [a^{l}_{\vec m}, a^{l, \dagger}_{\vec m}] = + 1 \
  . \nonumber \\
&&hw(4)_r: [n^r_{\vec m}, a^{r, \dagger}_{\vec m}] = - a^{r,
  \dagger}_{\vec m} \ , \nonumber \\
&&\qquad \qquad [n^r_{\vec m}, a^{r}_{\vec m}] = + a^{r}_{\vec m} \ ,
  \nonumber \\
&&\qquad \qquad [a^{r}_{\vec m}, a^{r, \dagger}_{\vec m}] = - 1 \ .
\ea
It is evident that hw(4)$_l$ (hw(4)$_r$) is isomorphic
(antiisomorphic, respectively) to hw(4). Since hw(4)$_l$ and
hw(4)$_r$ act in different spaces, the operators belonging to these
two different groups commute with each other. We write this somewhat
symbolically as
\be
\label{eq8}
[hw(4)_l, hw(4)_r] = 0 \ .
\ee

\subsection{Composite Algebra}
\label{com}

In order to explore the algebraic structure of Eq.~(\ref{eq3}), it is
not sufficient to list the basic algebras hw(4)$_l$ and hw(4)$_r$. It
is neccessary, in addition, to display the structure of the composite
algebra $C$ obtained by joining these two algebras. The composite
algebra $C$ consists of the elements
\be
C : \{ n^l_{\vec m}, n^r_{\vec m}, a^{r \dagger}_{\vec m} a^{l}_{\vec
m}, a^{l \dagger}_{\vec m} a^{r}_{\vec m} \} \ .
\label{eq9}
\ee
The elements of $C$ obey the commutation relations
\ba
\label{eq10}
&&[n^r_{\vec m}, a^{r \dagger}_{\vec m} a^{l}_{\vec m}] = - a^{r
\dagger}_{\vec m} a^{l}_{\vec m} \ , \nonumber \\
&&[n^r_{\vec m}, a^{l \dagger}_{\vec m} a^{r}_{\vec m}] = + a^{r}_{\vec
    m} a^{l \dagger}_{\vec m} \ , \nonumber \\
&&[n^l_{\vec m}, a^{r \dagger}_{\vec m} a^{l}_{\vec m}] = - a^{r
\dagger}_{\vec m} a^{l}_{\vec m} \ , \nonumber \\
&&[n^l_{\vec m}, a^{l \dagger}_{\vec m} a^{r}_{\vec m}] = + a^{r}_{\vec
    m} a^{l \dagger}_{\vec m} \ , \nonumber \\
&&[a^{l \dagger}_{\vec m} a^{r}_{\vec m}, a^{r \dagger}_{\vec m}
  a^{l}_{\vec m}] = - n^{r}_{\vec m} - n^{l}_{\vec m} \ , \nonumber \\
&&[n^{r}_{\vec m}, n^{l}_{\vec m}] = 0 \ .
\ea
These relations follow from Eqs.~(\ref{eq7}) and (\ref{eq8}). The
algebra $C$ is the direct sum U(1)$\oplus$SU(1,1) of two commuting
parts, a radical U(1) and a simple Lie algebra SU(1,1). Here
\be
\label{eq11}
U(1) : \{ n_{\vec m} = n^l_{\vec m} - n^r_{\vec m} \}
\ee
and
\be
\label{eq12}
SU(1,1) : \{ K^0_{\vec m} = \frac{1}{2} (n^{r}_{\vec m} + n^{l}_{\vec
  m}), K^{+}_{\vec m} = a^{l \dagger}_{\vec m} a^{r}_{\vec m},
  K^{-}_{\vec m} = a^{r \dagger}_{\vec m} a^{l}_{\vec m} \} \ .
\ee
The elements of SU(1,1) obey the relations
\ba
\label{eq13}
&&[K^{0}_{\vec m}, K^{\pm}_{\vec m}] = \pm K^{\pm}_{\vec m} \ ,
\nonumber \\
&&[K^{-}_{\vec m}, K^{+}_{\vec m}] = 2 K^{0}_{\vec m} \ .
\ea
We list the action of the four elements of $C$ on the projectors $
\Pi^N_{\vec m} = | N \rangle_{\vec m} {_{\vec m}} \langle N |$.
\ba
\label{eq14}
&&n_{\vec m} \Pi^N_{\vec m} = (-) \Pi^N_{\vec m} \ ,
\nonumber \\
&&K^{0}_{\vec m} \Pi^N_{\vec m} = \frac{1}{2}(2N + 1) \Pi^N_{\vec m}
\ , \nonumber \\
&&K^{+}_{\vec m} \Pi^N_{\vec m} = (N + 1) \Pi^{N+1}_{\vec m} \ ,
\nonumber \\
&&K^{-}_{\vec m} \Pi^N_{\vec m} = N \Pi^{N-1}_{\vec m} \ .
\ea
Since $\rho(t)$ is diagonal in energy representation, we are
concerned exclusively with the projectors $\Pi^N_{\vec m}$, i.e.,
with the eigenvalue $(-1)$ of the operator $n_{\vec m}$. Our
algebraic method is not confined to this case, however. This is
shown in the Appendix.

\subsection{The Structure of Eq.~(\ref{eq4})}
\label{str}

We define
\ba
\label{eq15}
&&\Gamma_{\vec m} = \sum_{{\vec n} \neq {\vec m}} \Gamma^{{\vec m},{\vec
    n}}_{{\vec n},{\vec m}} N_{\vec n} \ , \nonumber \\
&&\Gamma^{\vec m} = \sum_{{\vec n} \neq {\vec m}} \Gamma^{{\vec n},{\vec
    m}}_{{\vec m},{\vec n}} (N_{\vec n} + 1)
\ea
and use Eqs.~(\ref{eq11},\ref{eq12}) and (\ref{eq14}). The master
equation takes the form of a rate equation,
\be
\label{eq16}
\frac{{\rm d} \rho_A(t)}{{\rm d}t} =  \sum_{\vec m} \{ 2 \Gamma_{\vec
  m} K^+_{\vec m} + 2 \Gamma^{\vec m} K^-_{\vec m} - 2(\Gamma_{\vec
  m} + \Gamma^{\vec m}) K^0_{\vec m} - (\Gamma_{\vec m} - \Gamma^{\vec
  m}) \} \rho_A(t) \ .
\ee
Eq.~(\ref{eq16}) shows that aside from the mean values $N_{\vec m}$
contained in the rate coefficients $\Gamma^{\vec m}$ and $\Gamma_{\vec
  m}$, the time--dependence of $\rho_A(t)$ is given by a sum of terms
each referring to one of the orbital subspaces only. Moreover, each
such term is linear in the generators of SU(1,1) and is, thus,
integrable. This makes it possible to solve Eq.~(\ref{eq16})
algebraically.

We reduce the rate equation Eq.~(\ref{eq16}) by taking a partial
trace. We use a Hilbert--space representation in terms of products of
states $| N_{\vec m} \rangle$. We take the trace of $\rho_A(t)$ over
all states ${\vec n}$ with ${\vec n} \neq {\vec m}$ and denote the
result by $\rho_{\vec m}(t)$. From Eq.~(\ref{eq16}), we find
\be
\label{eq17}
\frac{{\rm d} \rho_{\vec m}(t)}{{\rm d}t} = [2 \Gamma_{\vec m}
K^+_{\vec m} + 2 \Gamma^{\vec m} K^-_{\vec m} - 2 (\Gamma_{\vec m} +
\Gamma^{\vec m}) K^0_{\vec m} - (\Gamma_{\vec m} - \Gamma^{\vec m})]
\rho_{\vec m}(t) \ .
\ee
The omission of the interaction term $H{\prime}_{A-A}$ from
Eq.~(\ref{eq17}) was justified in Refs.~\cite{pap01}. We recall that
$\rho_A(t)$ is diagonal in energy representation. It follows that
$\rho_{\vec m}$ can be written as a sum of the projectors
$\Pi^N_{\vec m}$,
\be
\label{eq18}
\rho_{\vec m}(t) = \sum_{N = 0}^{N_A} P^N_{\vec m} \Pi^N_{\vec m} \ .
\ee
We use this form in Eq.~(\ref{eq17}) and find for $N < N_A$
\be
\label{eq19}
\frac{{\rm d} P^N_{\vec m}(t)}{{\rm d}t} = 2 \Gamma_{\vec m} N
P^{N-1}_{\vec m} + 2 \Gamma^{\vec m} (N+1) P^{N+1}_{\vec m} - 2
(\Gamma_{\vec m} + \Gamma^{\vec m}) N P^N_{\vec m} - 2 \Gamma_{\vec m}
P^N_{\vec m} \ .
\ee
Eq.~(\ref{eq19}) is similar to the time--dependent Hartree--Fock
equation used in Nuclear Physics~\cite{rin80}. From
Eqs.~(\ref{eq17},\ref{eq18}) it follows that the mean particle number
$N_{\vec m}(t)$ in orbit ${\vec m}$ obeys the equation
\be
\label{eq20}
\frac{{\rm d} N_{\vec m}(t)}{{\rm d}t} = 2 \Gamma_{\vec m} (N_{\vec m}
+ 1) - 2 \Gamma^{\vec m} N_{\vec m} \ .
\ee
Using the definitions for the $\Gamma$'s in Eqs.~(\ref{eq15}) we see
that Eq.~(\ref{eq20}) coincides with Eq.~(39) of the second of
Refs.~\cite{pap01}. In particular, we have ${\rm d} \sum_{\vec m}
N_{\vec m} / {\rm d}t = 0$ so that particle number $N_A = \sum_{\vec m}
N_{\vec m}$ is conserved. This shows that the factorization assumption
used in Refs.~\cite{pap01} is equivalent to the mean--field
approximation.

\section{Thermodynamic Equilibrium: Stationary Solution}
\label{sta}

In thermodynamic equilibrium, the reduced density matrix becomes
independent of time, and the stationary solution of the rate equation
Eq.~(\ref{eq19}) obeys ${\rm d} \rho_A(t) / {\rm d}t = 0$ or
\be
\label{eq21}
\sum_{\vec m} \{ 2 \Gamma^{\vec m} K^+_{\vec m} + 2 \Gamma^{\vec m}
K^-_{\vec m} - 2 (\Gamma_{\vec m} + \Gamma^{\vec m}) K^0_{\vec m} -
(\Gamma_{\vec m} - \Gamma^{\vec m} \} \rho_A = 0 \ .
\ee
Inserting the expansion Eq.~(\ref{eq18}), we find the following set of
equations.
\ba
\label{eq22}
&&2 \Gamma_{\vec m} N P^{N-1}_{\vec m} + 2 \Gamma^{\vec m} (N+1)
P^{N+1}_{\vec m} - 2 (\Gamma_{\vec m} + \Gamma^{\vec m}) N P^N_{\vec m}
- 2 \Gamma_{\vec m} P^N_{\vec m} = 0 \ \ {\rm if} \ \ N < N_A \ ;
\nonumber \\
&&2 \Gamma_{\vec m} N_A P^{N_A-1}_{\vec m} - 2 (\Gamma_{\vec m} +
\Gamma^{\vec m}) N_A  P^{N_A}_{\vec m} - 2 \Gamma_{\vec m}
P^{N_A}_{\vec m} = 0 \ ; \nonumber \\
&&2 \Gamma_{\vec m}( N_A+1) P^{N_A}_{\vec m} = 0 \ .
\ea
It is easily seen that these equations only have the trivial solution
$P^{N}_{\vec m} = 0$ for all $N \leq N_A$. This fact shows that a
consistent non--trivial solution of the mean--field equations does not
exist if we impose on $\rho_A$ the constraint expressed by
Eq.~(\ref{eq18}). Therefore, we modify this equation by writing
\be
\label{eq23}
\rho_{\vec m}(t) = \sum_{N = 0}^{\infty} P^N_{\vec m} \Pi^N_{\vec m} \ .
\ee
All we now require is that the sum $\sum_{N = N_A}^{\infty}
P^N_{\vec m}$ be negligibly small. Then the first of Eqs.~(\ref{eq22})
applies for all $N$. The solution is
\be
\label{eq24}
P^N_{\vec m} = P^0_{\vec m} \ \chi^N \ \ {\rm where} \ \ \chi =
\frac{\Gamma_{\vec m}}{\Gamma^{\vec m}} \ .
\ee
From Eq.~(\ref{eq20}) we see that in the stationary case we always have
$\chi < 1$. The normalization condition yields $P^0_{\vec m} = (1
- \chi)$. The mean value $N_{\vec m}$ is given by $N_{\vec m} = \chi
/ (1 - \chi)$. Conversely, we may replace $\chi$ everywhere by
$N_{\vec m}/(N_{\vec m}+1)$. To discuss the validity of the mean--field
solution, we impose the constraint that $P^{N_A}_{\vec m}/P^{0}_{\vec m}
< \exp(-a)$. For $N_A \gg 1$, this yields $N_{\vec m}/N_A < 1/a$. This
shows that for the condensate (the only case where we expect
$N_{\vec m}$ to take sizeable values), the mean--field approximation
(as defined in the framework of this paper) begins to fail whenever the
ratio $N_{\vec m}/N_A$ grows beyond ten percent or so.

The $P^N_{\vec m} \sim \chi^N$ form a sequence which decreases
monotonically with increasing $N$. This behavior is in marked contrast
to that of the equilibrium solution found first by Scully~\cite{scu99}
and reproduced, in the present context, in Ref.~\cite{pap01}. Scully's
solution is not based upon the mean--field approximation. We believe
that it describes correctly the equilibrium behavior of the fully
developed condensate.

\section{Diagonalization of the Rate Operator}
\label{dia}

We turn to the time dependence of the solutions of the rate equation.
As a preparatory step, we determine in this Section the eigenvalues
and eigenvectors of the rate operator $\Gamma$, defined as the operator
which appears on the right--hand side of Eq.~(\ref{eq17}),
\be
\label{eq25}
\Gamma = [2 \Gamma_{\vec m} K^+_{\vec m} + 2 \Gamma^{\vec m}
K^-_{\vec m} - 2 (\Gamma_{\vec m} + \Gamma^{\vec m}) K^0_{\vec m} -
(\Gamma_{\vec m} - \Gamma^{\vec m})] \ .
\ee
For simplicity, we omit reference to the single--particle state
${\vec m}$, define
\be
\label{eq25a}
a = 2 \Gamma_{\vec m}, \ b = 2 \Gamma^{\vec m}
\ee
and have
\be
\label{eq26}
\Gamma = a K^+_{\vec m} + b K^-_{\vec m} - (a + b) K^0_{\vec m} -
\frac{1}{2} (a - b) \ .
\ee
We note that $\Gamma$ is not self--adjoint, and that
\be
\label{eq27}
\Gamma^{\dagger} = a K^-_{\vec m} + b K^+_{\vec m} - (a + b)
K^0_{\vec m} - \frac{1}{2} (a - b) \ .
\ee
The mean--field equation~(\ref{eq17}) takes the form
\be
\label{eq28}
\frac{{\rm d} \rho_{\vec m}(t)}{{\rm d}t} = \Gamma \rho_{\vec m} \ .
\ee
The equilibrium solution determined in section~\ref{sta} obviously
corresponds to the eigenvalue zero of the rate operator. The
eigenfunction is given by Eqs.~(\ref{eq23},\ref{eq24}). We now
construct the complete set of eigenvalues and eigenvectors of the
rate operator. A similar procedure has also been used in Quantum
Optics~\cite{bri93}.

To this end, it is useful to first look at a simpler problem. Given
the three components $J_1, J_2, J_3$ of spin which obey the cyclic
commutation relations $[J_1, J_2] = i J_3$, we ask for the
eigenvalues and eigenvectors of the operator
\be
\label{eq29}
{\cal O} = a J_2 + b J_3 \ .
\ee
The constants $a, b$ are real. Writing $a = \sqrt{a^2 + b^2} \ \sin
\theta$ and $b = \sqrt{a^2 + b^2} \ \cos \theta$, we have
\be
\label{eq30}
{\cal O} = \sqrt{a^2 + b^2} \ ( \sin \theta J_2 + \cos \theta J_3)
\ ,
\ee
and we recognize immediately that ${\cal O}$ can be diagonalized by
a rotation in the $(2, 3)$ plane by the angle $- \theta$, i.e., by
the unitary transformation $U = \exp( - i \theta J_1)$,
\be
\label{eq31}
\overline{{\cal O}} = U {\cal O} U^{\dagger} = \sqrt{a^2 + b^2} \
J_3 \ .
\ee
This result can be verified algebraically. The eigenvalues of
${\cal O}$ are proportional to those of the operator $J_3$, the
eigenvectors are obtained by applying the unitary transformation
$U^{\dagger}$ to the eigenvectors of $J_3$. The diagonalizing matrix
$U$ is not unique: Replacing $\theta$ by $\theta \pm \pi$ also
yields a diagonal form, with $J_3$ replaced by $(- J_3)$.

The form of the rate operator in Eq.~(\ref{eq26}) is somewhat more
complicated than that of the operator ${\cal O}$. Moreover, the
underlying group is not SU(2) but SU(1,1). Therefore, it takes a
similarity rather than a unitary transformation to diagonalize
$\Gamma$. Geometrically speaking, the vector $\Gamma$ is not
confined to the $(2,3)$ plane. This makes it necessary to define a
pair of similarity transformations,
\be
\label{eq32}
T_{-} = \exp(- \alpha_{-} K^{-}) \ , \ T_{+} = \exp(- \alpha_{+}
K^{+}) \ .
\ee
Eqs.~(\ref{eq13}) imply the identities
\ba
\label{eq32a}
&&\exp(- \alpha_{-} K^{-}) K^{+} \exp( \alpha_{-} K^{-}) = K^{+}
- 2 \alpha_{-} K^{0} - \alpha_{-}^2 K^{-} \ ; \nonumber \\
&&\exp(- \alpha_{-} K^{-}) K^{0} \exp( \alpha_{-} K^{-}) = K^{0}
- \alpha_{-} K^{-} \ ; \nonumber \\
&&\exp(- \alpha_{+} K^{+}) K^{-} \exp( \alpha_{+} K^{+}) = K^{-}
+ 2 \alpha_{+} K^{0} + \alpha_{+}^2 K^{+} \ ; \nonumber \\
&&\exp(- \alpha_{+} K^{+}) K^{0} \exp( \alpha_{+} K^{+}) = K^{0}
+ \alpha_{+} K^{+} \ .
\ea
It is easy to see that the product
\be
\label{eq33}
T = T_{-} T_{+}
\ee
diagonalizes $\Gamma$ (in the sense that $T \Gamma T^{-1}$ is
diagonal) provided the coefficients $\alpha_{-}$ and $\alpha_{+}$
obey the equations
\be
\label{eq34}
a + b \ \alpha_{+}^2 - (a + b) \ \alpha_{+} = 0 \ ; \ \ b \ ( 1
- 2 \alpha_{+} \alpha_{-} ) + ( a + b ) \ \alpha_{-} = 0 \ .
\ee
The transformed rate operator reads
\be
\label{eq35}
T \Gamma T^{-1} = [ 2 b \alpha_{+} - ( a + b ) ] \ K^{0} -
\frac{1}{2} \ (a - b) \ .
\ee
The eigenvectors  of $T \Gamma T^{-1}$ belonging to the eigenvalue
$-1$ of the operator $n_{\vec{m}}$, are those of $K^{0}$ and have
the form $\Pi^n$, with $n = 0, 1, 2, \ldots$. The right--hand
eigenvectors of the rate operator $\Gamma$ are correspondingly
given by $c_n T^{-1} \ \Pi^n$ with $c_n$ a normalization constant.

Eqs.~(\ref{eq34}) have two solutions, yielding two sets of
eigenvalues $\gamma_n$ for the rate operator. This fact is due to
the same type of ambiguity as found for the operator ${\cal O}$ in
the last but one paragraph. With $n = 0, 1, 2, \ldots$, the
solutions are
\ba
\label{eq36}
&& (a) \ \ \alpha_{+} = \chi \ , \ \alpha_{-} = \frac{1}{\chi - 1}
\ , \ \gamma_n = n (a - b) \ ; \nonumber \\
&& (b) \ \ \alpha_{+} = 1 \ , \ \alpha_{-} = \frac{1}{1 - \chi} \ ,
\ \gamma_n = (n + 1) (b - a) \ .
\ea
For the right--hand eigenvectors of $\Gamma$, this yields
\be
\label{eq37}
\rho_n = c_n T^{-1} \ \Pi^n = c_n \exp ( \alpha_{+} K^{+}) \exp
( \alpha_{-} K_{-} ) \ \Pi^n \ ,
\ee
where $c_n$ is a normalization constant. Eq.~(\ref{eq37}) shows that
the solutions $(b)$ of Eq.~(\ref{eq36}) are not admissible. Indeed,
we have $\exp ( \alpha_{+} K^{+} ) \ \Pi^0 = \sum_{N=0}^{\infty} \
\alpha_{+}^N \Pi^N$, and a corresponding result when the operator
$\exp ( \alpha_{+} K^{+} )$ acts on $\Pi^N$ with $N \geq 1$. For
solution $(a)$ of Eq.~(\ref{eq36}), where $\alpha_{+} = \chi < 1$,
the resulting series behaves as required while for solution $(b)$
where $\alpha_{+} = 1$, the coefficients multiplying $\Pi^N$ do
not decrease with decreasing $N$ and, thus, violate the condition
formulated in Section~\ref{sta} for the viability of the
mean--field solution. Solution $(a)$ does contain the zero
eigenvalue discussed in Section~\ref{sta}. Moreover, all non--zero
eigenvalues $\gamma_n$ of solution $(a)$ are negative, as is to be
expected on physical grounds.

Eq.~(\ref{eq35}) shows that the operator $T \Gamma T^{-1}$ is
self--adjoint and, thus, equal to $(T^{-1})^{\dagger}
\Gamma^{\dagger} T^{\dagger}$. This implies that $\Gamma^{\dagger}$
has the same eigenvalues as $\Gamma$ and is diagonalized by the
similarity transformation $(T^{-1})^{\dagger}$. The right--hand
eigenvectors of $\Gamma^{\dagger}$ are given by $c_n T^{\dagger} \
\Pi^n$. They coincide with the left--hand eigenvectors
$\tilde{\rho}_n$ of $\Gamma$. It is easy to show that the right--
and left--hand eigenvectors of $\Gamma$ form a biorthogonal
set~\cite{mor80},
\be
\label{eq38}
{\rm Tr} ( \tilde{\rho}_n \rho_m ) = \delta_{n m} \ .
\ee

\section{Time--Dependent Solutions of the Mean--Field Equations}
\label{tim}

In the fully time--dependent mean--field equations Eq.~(\ref{eq28}),
the rate operator $\Gamma$ depends upon time. This is because
$\Gamma$ is defined in terms of the coefficients $a$ and $b$ which
in turn depend on time via the mean occupation numbers $N_{\vec m}$,
see Eqs.(\ref{eq25a}) and (\ref{eq15}). We now show that given the
solutions $N_{\vec m}(t)$ of the rate equations Eq.~(\ref{eq20})
and, thus, the time dependence of the coefficients $a(t)$ and
$b(t)$, the solutions to the full equations~(\ref{eq28}) are
obtained by quadratures. We also show that these solutions tend
asymptotically towards the equilibrium solution found in
Section~\ref{sta}. As in Section~\ref{dia}, we suppress the label
for the single--particle state ${\vec m}$. We retain the symbol
$\chi$ introduced in Section~\ref{sta} to denote the asymptotic or
equilibrium value of the ratio $a/b$.

We assume that at time $t = 0$, we are given the initial density
matrix
\be
\label{eq38a}
\rho(0) = \sum_{N=0}^{\infty} P^N \Pi^N
\ee
with the proviso that for $N > N_A$, the coefficients $P^N$ are
negligible. To solve Eq.~(\ref{eq28}), we define the
time--dependent similarity transformation
\be
\label{eq39}
\overline{\rho(t)} = T(t) \ \rho(t) = \exp[ - \alpha_{-}(t) \ K^{-}
] \ \exp[ - \alpha_{+}(t) \ K^{+} ] \ \rho(t) \ .
\ee
The time--dependent coefficients $\alpha_{\pm}(t)$ are solutions of
the differential equations
\be
\label{eq40}
\frac{{\rm d} \alpha_{+}(t)}{{\rm d}t} = a + b \alpha_{+}^2 - (a
+ b) \ \alpha_{+} \ ; \ \frac{{\rm d} \alpha_{-}(t)}{{\rm d}t} = b
\ (1 - 2 \alpha_{+} \alpha_{-}) + (a + b) \ \alpha_{-} \ .
\ee
Comparison with Eqs.~(\ref{eq33},\ref{eq34}) shows that $T(t)$
provides the time--dependent generalization of the similarity
transformation $T$ used in Section~\ref{dia}. As initial conditions
for the differential equations~(\ref{eq40}), we choose
$\alpha_{\pm}(0) = 0$ or $\overline{\rho(0)} = \rho(0)$. Using the
identities Eq.~(\ref{eq32a}), we see that the transformed density
matrix $\overline{\rho(t)}$ obeys
\be
\label{eq41}
\frac{{\rm d} \overline{\rho(t)}}{{\rm d}t} = [ \gamma(t) \ K^0 -
\frac{1}{2} (a - b) ] \ \overline{\rho(t)} \ ,
\ee
where
\be
\label{eq42}
\gamma(t) = 2 b \alpha_{+} - (a + b) \ .
\ee
Integrating this equation and using the relation inverse to
Eq.~(\ref{eq39}), we obtain
\be
\label{eq43}
\rho(t) = \exp[ \alpha_{+}(t) K^{+} ] \ \exp[ \alpha_{-}(t) K^{-}
] \ \exp \biggl( \int_0^t {\rm d}\tau [ \gamma(\tau) K^0 - \frac{1}{2}
(a(\tau) - b(\tau)) ] \biggr) \ \rho(0) \ .
\ee
We have accomplished our first goal and shown that given the
time dependence of the coefficients $a$ and $b$, the time
dependence of the density matrix is obtained by quadratures.

We now show that $\rho(t)$ as given by Eq.~(\ref{eq43}) tends
asymptotically ($t \rightarrow \infty$) to the equilibrium solution
established in Section~\ref{sta}. To this end, we first investigate
the asymptotic behavior of $\alpha_{\pm}(t)$. The first of
Eqs.~(\ref{eq40}) reads ${\rm d} \alpha_{+}(t)/{\rm d}t = b
(\alpha_{+} - a/b) (\alpha_{+} - 1)$. We recall that $a > 0, b > 0$
and distinguish two cases. (a) We have $a/b < 1$ for all times $t$.
Then, ${\rm d} \alpha_{+}(t)/{\rm d}t > 0 (< 0)$ if $0 <
\alpha_{+}(t) < a/b$ (if $a/b < \alpha_{+}(t) < 1$, respectively).
With the initial condition $\alpha_{+} = 0$, we see that $\alpha_{+}$
approaches the value $a(\infty)/b(\infty) = \chi < 1$ asymptotically
from below. This implies $\gamma(t) \rightarrow (a - b) < 0$ as $t
\rightarrow \infty$. (b) We have $a/b > 1$ for some time interval,
although $a/b \rightarrow \chi < 1$ must hold asymptotically. As
long as $a/b > 1$, the initial condition $\alpha_{+} = 0$ implies
$\alpha_{+}(t) < 1$. As soon as $a/b$ intersects the value unity
from above, $\alpha_{+}(t)$ has a negative (positive) slope if it
is larger (smaller) than $a/b$. In either case, $\alpha_{+}$
again tends toward $\chi < 1$ asymptotically, and $\gamma$
attains the asymptotic value $a(\infty) - b(\infty) < 0$ as
in case (a). To study the asymptotic behavior of $\alpha_{-}$,
we define $f(t) = \alpha_{-} \ \exp[\int_0^t {\rm d}\tau
\gamma(\tau)]$. The time derivative of $f(t)$ is given by $b
\exp[ \int_0^t {\rm d}\tau \gamma(\tau) ]$. Since $b$ is
bounded and $\gamma$ negative for large $t$, this derivative
tends to zero and, hence, $f$ towards a constant $f_0$. This
implies that $\alpha_{-}$ diverges asymptotically.

We turn to the asymptotic behavior of $\rho(t)$. We use
Eqs.~(\ref{eq38a}) and (\ref{eq14}) to write
\ba
\label{eq44}
&&\rho(t) = \exp[ \alpha_{+}(t) K^{+} ] \ \exp [ (1/2)
\int_0^{t} {\rm d}\tau (\gamma(\tau) - (a(\tau) - b(\tau) )]
\nonumber \\
&&\qquad \qquad \times \sum_{N=0}^{\infty} f(t)^N P^N(0)
\sum_{p=0}^N {N \choose p} \ (\alpha_{-})^{-p} \ \Pi^p \ .
\ea
With the help of the asymptotic behavior of $f(t), \alpha_{+}$,
and $\alpha_{-}$ established in the previous paragraph, this
becomes asymptotically
\be
\label{eq45}
\rho(t) \rightarrow \exp[ \chi K^{+} ] \ \exp [ (1/2)
\int_0^{\infty} {\rm d}\tau (\gamma(\tau) - (a(\tau) - b(\tau)
)] ( \sum_{N=0}^{\infty} f(0)^N P^N(0)) \ \Pi^0 \ .
\ee
As $\gamma$ tends asymptotically towards $(a(\infty) -
b(\infty))$, the integral in the exponent exists. Actually we
ought to show that the difference $\gamma - (a - b)$ vanishes
asymptotically at least as strongly as $t^{-2}$. This we have
not done. We know, however, from Eq.~(\ref{eq20}) that both
${\rm tr} \ \rho$ and the mean value $N_{\vec m}$ of the number
operator remain finite for $t \rightarrow \infty$. Therefore,
the normalization constant of our solution cannot diverge
asymptotically. Together with $( \sum_{N=0}^{\infty} f_0^N
P^N(0))$, the exponential can be lumped into a new constant
$C$. This yields
\be
\label{eq46}
\rho(t) \rightarrow C \exp[ \chi K^{+} ] \ \Pi^0 \ .
\ee
This shows that asymptotically, $\rho$ becomes proportional to
the equilibrium solution found in Section~\ref{sta}.

\section{Summary and Conclusions}
\label{sum}

We have used the mean--field approximation to simplify the
master equation for sympathetic cooling of Bosons. We have
shown that the factorization assumption of Refs.~\cite{pap01} is
equivalent to this approximation. Studying the stationary or
equilibrium solution of the resulting master equation, we have
shown that the mean--field approximation to sympathetic cooling
is expected to break down whenever the fraction of condensate
Bosons exceeds ten percent or so of the total Boson number in
the cooled gas. This conclusion is supported by the observation
that the equilibrium solution for the one--body density matrix
differs markedly from the form derived~\cite{scu99,pap01} for
the same quantity for the case of a fully developed condensate.
Using group--theoretical methods, we have solved the
time--dependent master equation for the one--body density
matrix. Given the time--dependence of the mean single--particle
occupation numbers, the solution is obtained by quadratures. It
tends asymptotically $(t \rightarrow \infty)$ towards the
equilibrium solution. For the time--dependence of the fully
developed condensate, quantum fluctuations are important. These
can probably be calculated along the lines described in Section
V of the second of Refs.\cite{pap01}.

We expect that the mean--field approximation will play an
important role in studies of sympathetic cooling. In this paper,
we believe to have given a comprehensive theoretical treatment of
this approximation.

{\bf Acknowledgment}. This work was supported in part by the
Max--Planck--Gesellschaft (MPG) and by the National Natural Science
Foundation of China. WSJ is grateful to the Max--Planck--Institute for
Nuclear Physics in Heidelberg for its hospitality. ANS acknowledges the
financial support given by 
FAPESP (Funda\c{c}\~ao de Amparo a Pesquisa do Estado de S\~ao Paulo). MCN acknowledges Conselho Nacional de Desenvolimento
Cient\'{\i}fico e Tecnol\'ogico (CNPq), the Max Planck Institute of Nuclear 
Physics at Heidelberg and the Humboldt Foundation.
\appendix

\section{General Approach to Equilibrium}

We remove the restriction to the eigenvalues $-1$ of the operator
$n_{\vec{m}}$ and define
\be
\Pi^{n,k}_{\vec{m}}=|n \rangle_{\vec{m} \vec{m}}\langle k| \ .
\ee
Eqs.~(\ref{eq14}) are replaced by 
\ba
\label{eqA2}
&&n_{\vec m} \Pi^{n,k}_{\vec m} = (n-k-1) \Pi^{n,k}_{\vec m} \ ,
\nonumber \\
&&K^{0}_{\vec m} \Pi^{n,k}_{\vec m} = \frac{1}{2}(n+k+1)
\Pi^{n,k}_{\vec m} \ , \nonumber \\
&&K^{+}_{\vec m} \Pi^{n,k}_{\vec m} = \sqrt{(n+1)(k+1)} \; 
\Pi^{n+1,k+1}_{\vec m} \ ,
\nonumber \\
&&K^{-}_{\vec m} \Pi^{n,k}_{\vec m} = \sqrt{nk} \; 
\Pi^{n-1,k-1}_{\vec m} \ .
\ea
The eigenvalues $\gamma_{n,k}$ of the transformed operator in
Eq.~(\ref{eq35}) are given by
\be
\gamma_{n,k} = \frac{1}{2}(a-b)(n+k+1)-\frac{1}{2}(a-b) \ ,
\ee
where $n$ and $k$ run over non--negative integers and the eigenvectors
are $c_{n,k} \Pi^{n,k}$. The right--hand eigenvectors of $\Gamma$ are
\be
\rho_{n,k} = c_{n,k}T^{-1}\Pi^{n,k} \ .
\ee
The approach to equilibrium can also be generalized by considering
\be
\label{eqA5}
\rho(t) = \sum_{n=0}^{\infty} \sum_{k=0}^{\infty} P^{n,k}(0)
\Pi^{n,k}(t)
\ee
where $P^{n,k}$ is defined by the initial condition. In analogy with 
Eq.~(\ref{eq44}), Eq.~(\ref{eqA5}) can be rewriten as
\ba
\rho(t) = & \exp [\alpha_+ K^{+}] \exp[(\frac{1}{2}) \int_0^{t} d\tau
(\gamma (\tau)-(a(\tau)-b(\tau))] \exp [\frac{n}{2}\int_0^t dt'
\gamma(t')] \nonumber \\
  & \nonumber \\
 & \times \sum_{n,k=0}^{\infty} f^k(t) P^{n,k}(0) \sum_{p=0}^k
\frac{(\alpha_{-})^{-p}}{(k-p)!} \sqrt{ \frac{n!k!}{(n-k+p)!p!}}
\; |n-k+p \rangle \langle p| \ .
\ea
We have assumed here $k<n$ for simplicity. Given the asymptotic limit
of the time--dependent function in the equation above one can
immediately conclude that as $t \rightarrow \infty$, only the term
corresponding to $n=k=0$ will survive, yielding the equilibrium state.


\begin{thebibliography}{99}

\bibitem{lew95}M. Lewenstein, J. I. Cirac, and P. Zoller,
  Phys. Rev. {\bf A 51} (1995) 4617.
\bibitem{pap01}T. Papenbrock, A. N. Salgueiro, and
  H. A. Weidenm\"uller, cond-mat/0106392 and cond-mat/01006418.
\bibitem{scu99}M. O. Scully, Phys. Rev. Lett. {\bf 82} (1999) 3927.
\bibitem{koc00}V. V. Kocharovsky, M. O. Scully, S.-Y. Zhu, and
  M. S. Zubairy, Phys. Rev. {\bf A 61} (2000) 023609.
\bibitem{wan93}S. J. Wang, F. L. Li, and A. Weiguny, Phys. Lett. {\bf
  A 180} (1993) 189.
\bibitem{wan89}S. J. Wang, J. M. Cao, and A. Weiguny, Phys. Rev. {\bf
  A 40} (1989) 1225.
\bibitem{rin80}P. Ring and P. Schuck, {\it The Nuclear Many--Body
  Problem}, Springer--Verlag, New York (1980).
\bibitem{bri93}H.-J. Briegel and B.-G. Englert, Phys. Rev. {\bf A 47}
  (1993) 3311.
\bibitem{mor80}P. Morse and H. Feshbach, {\it Methods of Theoretical
  Physics}, Wiley, New York (1980).

\end{thebibliography}
\end{document}